\documentstyle[12pt,wrapfig]{article}
\newcommand{\od}{\stackrel{\cdot}}
\newcommand{\td}{\stackrel{\cdot\cdot}}
\newcommand{\fd}{\stackrel{\cdot\cdot\cdot}}
\begin{document}
\begin{center}
{\Large Lienard-Wiechert potential and synchrotron radiation of a
relativistic spinning particle in the pseudoclassical theory}\\

{\large   Arakelyan S.A.,  Grigoryan
G.V.}\raisebox{.8ex}{$\star$},  {\large Grigoryan R.P.}
\raisebox{.8ex}{$\star\star$}.\footnote{ Partially supported  by
the grants INTAS 96-538,  INTAS 93-1038 and INTAS-RFBR 95-0829.}\\
{\em Yerevan Physics Institute,  Republic of Armenia}\\
\raisebox{.8ex}{$\star$}{E-mail:gagri@lx2.yerphi.am}\\
\raisebox{.8ex}{$\star\star$}{E-mail:rogri@lx2.yerphi.am}\\
\vspace{5mm}

\end{center}
\centerline{{\bf{Abstract}}}
 
Lienard-Wiechert potentials  of the relativistic spinning
particle with anomalous magnetic moment   in pseudoclassical
theory are constructed. General expressions for the
Lienard-Wiechert potentials are used for investigation of some
specific cases of the motion of the spinning particle. In
particular the spin dependence of the intensity of the
synchrotron radiation of the transversely polarized particle
performing uniform circular motion is considered. When the
particle moves in the external homogeneous magnetic field
the obtained formulae coincide with those known from quantum
theory of radiation. The dependence of the polarization of the
synchrotron radiation on the spin of the particle is
investigated.\\




{\bf1.}The introduction of Grassmann variables into the theory
allowed to construct theories of the point particles,  which
describe the spin of the particle already at the
classical level \cite{BM1, Casa}.
These theories are known as pseudoclassical theories and
are classical gauge theories with constraints. Methods for
quantization of these theories are presented in \cite{D1,  GTY,
HTQ}. The possibility of consistent description of the spin in
external fields using
Grassmann variables already at the classical level allows to
investigate spin effects in a number of physical processes using
classical equations of motion.
Note that  description of the spin in classical theories without
Grassmann variables is also possible,  however these theories  do not
give after quantization the accepted quantum relativistic
Dirac particle (see the review \cite{AF} and references therein).

In this paper in the pseudoclassical approach the
Lienard-Wiechert potentials for a charged  relativistic spinning
particle with anomalous magnetic moment (AMM) interacting with
the external electromagnetic field are obtained (a self-contained
pseudoclassical theory of the interaction of the spinning
particle with AMM with the external electromagnetic  field was
constructed in \cite{AZh,  GS,  GG5}).
Using these expressions for the Lienard--Wiechert potentials
certain cases of the charged particle movement are considered.
In particular the general expressions for the  dependences of
the intensity and the polarization of the synchrotron
radiation  on the polarization of the particle  for the
transversely polarized particle performing a uniform circular
motion are obtained. Investigations were carried out in
quadratic approximation in the Grassman variables,  or,  which is
the same,  to the first order in the spin.
 In this approximation,  in the
case of the movement in the homogeneous magnetic field,  the
obtained formulae (when the dependence of the effective mass of
the particle on the strengt of the magnetic field is taken into
account) coincide with similar formulae obtained in the classical
radiation theory \cite{TBMU1},  as well as with those in quantum
theory \cite{STSY}.

It is important to stress,  that within our approach the
evaluation of the formulae,  describing the spin dependence of the
measurable characteristics of the relativistic particle radiation
is much easier,  than in quantum electrodynamics.

Note,  that these potentials were constructed  also for
investigation of the problem of (classical) renormalization of the
pseudoclassical equations of motion of the  electron with
radiation reaction taken into account. The results of these
investigations will be reported in the separate paper.

This paper is organized as follows.
In section 2 we describe the Lagrangian of the pseudoclassical theory of
the spinning particle with AMM interacting with the
electromagnetic field. The equations of motion and fixation of
the gauges are written down,  additional gauge fixing conditions
are imposed  and general expressions for the Lienard--Wiechert potentials
are given. In section 3 the case of the particle at rest with
 precessing spin is considered. In section 4 the expressions for
 the synchrotron radiation field strengths for the relativistic
transversely polarized  spinning particle performing a uniform
circular motion  are deduced and the
dependences of the intensity and the polarization of the
synchrotron radiation on the polarization of the particle are
obtained. The special case of the movement in the  external
homogeneous magnetic field is considered in details.\\


{\bf2.} Thus,  consider  the action of the theory,  describing the
interaction of the relativistic spinning particle with anomalous
magnetic moment (AMM) with electromagnetic field in the space
time dimensions $D=4$ \cite{AZh,  GS,  GG5}):
\begin{eqnarray}
\label{Action}
 S &=& \frac{1}{2}\int
d\tau\left[\frac{\left(\dot x^\mu\right)^2}
{e}+em^2-i\left(\xi_\mu\dot \xi^\mu- \xi_5\dot
 \xi_5\right) +\right.\nonumber\\
&& +2g\dot x^\mu A_\mu+igMe
 F_{\mu\nu}\xi^\mu\xi^\nu
 -4iGF_{\mu\nu}\dot x^\mu\xi^\nu\xi_5-\\
&&\left.- i\chi \bigl(\frac{\xi_\mu\dot x^\mu}{e}-m\xi_5
-iGF_{\mu\nu}\xi^\mu\xi^\nu\xi_5\bigr)-
eG^2(F_{\mu\nu}\xi^\mu\xi^\nu)^2\right]+\frac{1}{4}\int
d^4zF_{\mu\nu}(z)F^{\mu\nu}(z).\nonumber
\end{eqnarray}

Here $ x^\mu(\tau)$  is particle coordinate,  ,  $\xi^\mu(\tau)$
are Grassmann  variables,  describing spin degrees of  freedom,
$\xi_5(\tau),   \chi$ and $e$  are additional fields ($e$ is an
even element,   $\xi_5$ and $\chi$ are odd elements  of Grassmann
algebra) ,    $A^\mu$   is
the  vector-potential of the electromagnetic field,  $F_{\mu\nu} =
\partial_\mu A_\nu - \partial_\nu A_\mu$; $g$ is the charge of
the  particle,     $(-G)$- the anomalous magnetic moment,
$M=1-2Gm/g$  is the total magnetic moment of the particle in Bohr
magnetons;   the overdote denotes the differentiation over $\tau$
along the  trajectory; the derivatives over  Grassmann variables
are left.

The theory  (\ref{Action})  has three   gauge
symmetries: reparametrization and supersymmetry \cite{GS},    and
also the  $U(1)$ gauge symmetry of electromagnetic interactions.
They correspond to the presence of three primary constraints of
the first class in the lagrangian formulation of the theory.
Fixation of the latter will be accomplished below.

To obtain the expression for Lienard-Wiechert potentials we'll
write the equation for the electromagnetic field $A_\mu$ and also
for the fields $e$, $\xi_5$ and $\chi$:
\begin{eqnarray}
\label{A2}
\frac{\delta S}{\delta
A_\lambda(y)}&=&-\partial_\mu F^{\mu\lambda}(y)+g\int
d\tau\dot{x}^\lambda\delta(x(\tau)-y)
-2iG\int d\tau (\dot{x}^\mu\xi^\lambda-\dot{x}^\lambda\xi^\mu)\xi_5
\partial_\mu\delta(x(\tau)-y)+\nonumber\\
&+&igM\int d\tau e\xi^\mu\xi^\lambda \partial_\mu\delta(x(\tau)-y)-
2G^2\int d\tau
eF_{\sigma\rho}\xi^\sigma\xi^\rho\xi^\mu\xi^\lambda
\partial_\mu\delta(x(\tau)-y)- \nonumber\\
&-&G\int  d\tau \chi \xi^\mu\xi^\lambda\xi_5
\partial_\mu\delta(x(\tau)-y)
=0, \\
&&\nonumber\\
\label{A3}
\frac{\delta S}{\delta
e}&=&-\frac{\dot{x}^2}{e^2}+m^2+igMF_{\mu\nu}\xi^\mu\xi^\nu
-G^2\left(F_{\mu\nu}\xi^\mu\xi^\nu\right)^2+i\chi
\frac{\xi_\mu\dot{x}^\mu}{e^2}=0, \\
&&\nonumber\\
\label{A4}
\frac{1}{i}\frac{\delta S}{\delta \chi}&=&\frac{\xi_\mu
\dot{x}^\mu}{e}-m\xi_5-iGF_{\mu\nu}\xi^\mu\xi^\nu\xi_5=0.\\
\label{A4B}
\frac{1}{i}\frac{\delta S}{\delta \xi_5}&=&2\od{\xi}_5+4GF_{\mu\nu}\od{x}^\mu\xi^\nu-
\chi\left(m+ iGF_{\mu\nu}\xi^\mu\xi^\nu\right)=0.
\end{eqnarray}

Now we will fix  one of the gauges of  the theory  by
imposing the condition
\begin{equation}
\label{A20}
\od {x}{\!}^\mu \xi_\mu=0,
\end{equation}
which due to relation (\ref{A4}) is equivalent to
$\xi_5=0$.
Using the latter equality from equations
(\ref{A3}) and (\ref{A4B}) we find the expressions for the
auxiliary fields $e$ and $\chi$:
\begin{eqnarray}
\label{A5}
e&=&-\sqrt{\frac{\dot{x}^2}{m^2+igMF_{\mu\nu}\xi^\mu\xi^\nu-
G^2\left(F_{\mu\nu}\xi^\mu\xi^\nu\right)^2}}=\nonumber\\
&=&-\frac{\sqrt{\dot{x}^2}}{m}\left[1-
\frac{igM}{2m^2}F_{\mu\nu}\xi^\mu\xi^\nu+
\left(\frac{G^2}{2m^2}
-\frac{3}{8}\left(\frac{gM}{m^2}\right)^2\right)
(F_{\mu\nu}\xi^\mu\xi^\nu)^2\right],\\
&&\nonumber\\
\label{A6}
\chi&=&\frac{4GF_{\mu\nu}\dot{x}^\mu\xi^\nu}{
m+iGF_{\mu\nu}\xi^\mu\xi^\nu}=\frac{4GF_{\mu\nu}\dot{x}^\mu\xi^\nu}{
m}(1-\frac{iG}{m}F_{\mu\nu}\xi^\mu\xi^\nu)\nonumber
\end{eqnarray}
(the choice of the sign of $e$ in the expression (\ref{A5}) is
dictated by the positivity of the energy in nonrelativistic
limit).

The expansion of the  right  hand  side  of  these
expressions in powers  of
$\xi_\mu$ is carried out with taking into account the fact  that
the powers of $\xi_\mu$ higher than four are equal to  zero.
Substituting  in  (\ref{A2})  $e$     by   the
expressions (\ref{A5}) and inserting    $\xi_5=0$ we get the  following
equation for the field $A_\mu$:
\begin{eqnarray}
\label{A7}
&&\left(\partial _\mu \partial ^\mu A^\lambda -\partial ^\lambda
\partial _\mu A^\mu  \right)(y)+\left(\frac{2G^2}{m}-
\frac{g^2M^2}{2m^3}\right) \frac{\partial }{\partial y^\mu }\int
d\tau  \delta (x-y)\sqrt{\dot x^2}F_{\sigma \rho }\xi ^\sigma \xi
^\rho \xi ^\mu \xi ^\lambda-\nonumber\\
&-&\frac{G}{m^2}(gM-2Gm)\frac{\partial}{\partial
y^\mu}\int d\tau\frac{\delta(x(\tau)-y)}{\sqrt{\dot
x^2}}\left[\dot x^\mu \xi^\lambda -\dot x^\lambda \xi ^\mu \right]
\dot x^\nu \xi_\nu F_{\sigma\rho}\xi ^\sigma \xi ^\rho =j^\lambda (y),
\end{eqnarray}
where  $j^\lambda (y)$ is given by the expression
\begin{eqnarray}           
\label{A8}
j^\lambda(y)&=& g\int d\tau\dot x^\lambda\delta(x(\tau)-y)
 -2iG\frac{\partial }{\partial y^\mu }
\int d\tau\frac{\delta(x(\tau)-y)}{\sqrt{\dot x^2}}
\left[\dot x^\mu \xi ^\lambda -\dot x^\lambda \xi ^\mu \right]
\xi_\nu\dot x^\nu+\nonumber\\
&+&\frac{igM}{m}\frac{\partial }{\partial
y^\mu }\int  d\tau\delta (x-y)\sqrt{\dot x^2} \xi^\mu \xi
^\lambda.
\end{eqnarray}
Further simplification of this equation is achieved  by  the
following. Let us fix two remaining  gauges  of
the theory. To do this we will choose the gauge condition for
the field $A_\mu$ in the form  $\partial_\mu  A^\mu=0$,   and
will identify the parameter $\tau$ with the proper  time  of
the  particle,   which  means  fixation  of  the  $U(1)$  and
reparametrization degrees of freedom correspondingly. Then we
get ${\dot x}^2=1$. Also we will consider only  those  terms
of  the  equation  (\ref{A7}),   which  contain   powers   of
$\xi_\mu$  no  higher  than  two.   This   limitations   are
equivalent  to  quasiclassical  approximation  since   after
quantization of the theory the terms containing second orders of
$\xi$  will  be  proportional  to  $\hbar$  (the  terms,
containing the fourth power of $\xi_\mu$ will give rise  to
corrections  of  the  order  of  ${\hbar}^2$ \cite{BM1, Casa}).
Taking  into account all this,  we can rewrite the equation
(\ref{A7})  in the form:
\begin{equation}
\label{A9}
\Box A^\mu=j^\mu,
\end{equation}
where $j^\mu$ is given by the expression (\ref{A8}).  It  is
convenient to represent the current vector $j^\mu$ as a  sum
of two terms:
\begin{equation}
\label{A10}
j^\mu(y)=g\int d\tau\dot x^\mu\delta(x(\tau)-y)+
\frac{\partial}{\partial y^\nu}\int
d\tau\delta(x(\tau)-y)p^{\nu\mu}(\tau),
\end{equation}
where
\begin{equation}
\label{A11}
p^{\mu\nu}=\frac{i}{m}\left\{gM\xi^\mu\xi^\nu-2Gm\left[\dot
x^\mu\xi^\nu-\dot x^\nu\xi^\mu\right](\xi\dot x)\right\},
\quad p^{\mu\nu}=-p^{\nu\mu}.
\end{equation}
The first summand in (\ref{A10}) corresponds to the  current
of the charged particle without the dipole moment. The second
term corresponds to the contribution to the current  of  the
dipole moment  of the spinning particle.

It is easy to see, that due to (\ref{A20}) the electric dipole
moment of the particle $q_\mu$,  which is connected with
$p_{\mu\nu}$ by the relation $q_\mu=p_{\mu\nu}\od {x}{\!}^\nu$
(see e.g. \cite{EL}),  is equal to zero:
\begin{equation}
\label{A21}
q_\mu=\frac{ig}{m}\xi_\mu(\od{x} \xi)=0,
\end{equation}
as it should be for a point particle.

     We  will  show  now,   that  from  the   pseudoclassical
expression (\ref{A11}) for $p^{\mu\nu}$  follows,   that,  as
expected,   the total magnetic moment of the particle is directed
along  the spin of the particle. To do  this  we  will use the
expression,  connecting the vector of the magnetic  moment  of  the particle
$m_\mu$  with  the  tensor  of  the  dipole  moment $p^{\mu\nu}$
\cite{EL}:
\begin{equation}
\label{A12}
m_\mu=\frac{1}{2}\varepsilon_{\mu\nu\lambda\sigma}\od{x}{\!}^\nu
p^{\lambda\sigma}.
\end{equation}

Taking into account,  that in the gauge $\chi=0\, $,  $(\od{x}\!\xi)=0$
we have $\od{x}{\!}^\mu={e \cal
P}^\mu$ \cite{GG5} ( ${\cal P}_\mu=({\cal E},  \vec{{\cal P}})$ is
4-momentum of the particle
),  which using  (\ref{A5}) and dropping terms proportional to
$\xi^4$,  takes the form (for $\dot{x}^2=1$)
\begin{equation}
\label{A12A}
\od{x}{\!}^\mu=-{\cal P}^\mu/m_{\rm eff},  \quad m_{\rm
eff}=m+\frac{igM}{2m}F^{\mu\nu}\xi_\mu\xi_\nu.
\end{equation}
The analog of the relation between the 4-momentum of the
particle and its mass in this case has the form of $\displaystyle
{\cal P}_\mu^2- m_{\rm eff}^2+ {\rm O}(\xi)^4=0$,  while the sign
$"-"$ in (\ref{A12A}) is connected with the definition of the
generalized momentum,  conjugated to the coordinate of the
particle.

Inserting the expressions (\ref{A11}) and (\ref{A12A}) into (\ref{A12}) and
dropping the terms proportional to $\xi^4$,  we get
\begin{equation}
\label{A13}
m_\mu=-\frac{i}{2}\frac{gM}{m}\varepsilon_{\mu\nu\lambda\sigma}\frac{{
\cal P}^\nu}{m}\xi^\lambda\xi^\sigma=
\frac{gM}{m}\frac{W_\mu}{m}=M\frac{g}{2m}a_\mu.
\end{equation}
Here $W^\mu$ denotes the pseudoclassical analog of the Pauli-
Lubansky vector and is defined as \cite{GG5}
\begin{equation}
\label{A14}
W_\mu=\frac{-i}{2}\varepsilon_{\mu\nu\lambda\sigma}{\cal P}{\!}^\nu
\xi^\lambda\xi^\sigma,
\end{equation}
$a_\mu/2$ is the relativistic generalization of the
pseudoclassical spin vector (after canonical quantization the
vector $a_\mu$ becomes a polarization four vector of the spinning
particle). From (\ref{A13}) follows the proportionality of the
magnetic moment of the particle to the polarization vector with
the coefficient equal to the total magnetic moment of the
particle.

From the identity
\begin{equation}
\label{A22}
p_{\mu\nu}=\left(\delta_\mu^\alpha\delta_\nu^\beta-
\delta_\mu^\beta\delta_\nu^\alpha\right)q_\alpha\od
{x}{\!}_\beta +\varepsilon_{\mu\nu\lambda\sigma}m^\lambda\od
{x}{\!}^\sigma
\end{equation}
with accounting for (\ref{A21}) and  (\ref{A13}) follows the
equation
\begin{equation}
\label{A23}
p_{\mu\nu}=\frac{gM}{m^2}
\varepsilon_{\mu\nu\lambda\sigma}W^\lambda\od {x}{\!}^\sigma,
\end{equation}
which will be useful below.

The solution of the equation   (\ref{A9}) (with the current
represented in the form of (\ref{A10})) in terms of the retarded
fields was found in \cite{EL} and is given by the expressions
\begin{eqnarray}
\label{A15}
A^\mu(y)=A_{\rm
ret}^\mu(y)&=&\frac{1}{4\pi}\left[\frac{g\od{x}{\!}^\mu}{\rho}\right]
_{\tau=\tau_r}-\\
&-&\frac{1}{4\pi}\left\{\frac{1}{\rho}\left(\od{p}{\!}^{\mu\nu}k_\nu-
p^{\mu\nu}(\td{x}k)k_\nu\right)+\frac{1}{\rho^2}\left(p^{\mu\nu}k_\nu
-p^{\mu\nu}\od{x_\nu}\right)
\right\}_{\tau=\tau_r}, \nonumber
\end{eqnarray}
\begin{eqnarray}
\label{A16}
F_{{\rm ret}}^{\mu\nu}(y)&=&\frac{g}{4\pi}\left[\frac{2}{
\rho}\left(k^{[\mu}\td{x}{\!}^{\nu]}-
(\td{x}k)k^{[\mu}\od{x}{\!}^{\nu]}\right)
+\frac{2}{\rho^2}k^{[\mu}\od{x}{\!}^{\nu]}
\right]_{\tau=\tau_r}+\\
&+&
\frac{1}{4\pi}\left\{\frac{2P_1^{\left[\mu\nu\right]}}{\rho}+
\frac{2Q_1^{\left[\mu\nu\right]}}{\rho^2}+
\frac{2R_1^{\left[\mu\nu\right]}}{\rho^3}
\right\}_{\tau=\tau_r}\nonumber
\end{eqnarray}
where  $R^\mu\equiv y^\mu-x^\mu(\tau) $,  $R^\mu R_\mu=0$,
$\rho = \od{x}{\!\!}^\nu R_\nu$,  $k_\mu=R_\mu/\rho$; all
quantities in (\ref{A15}),  (\ref{A16}) are taken in the retarded time
 $\tau_r$ defined by the equation $t_0=t_0(\tau_r)=t-R(t_0)$
\footnote{Here and below square brackets
$[ \dots ]$ denote complete antisymmetrization:e.g.,

\centerline{$A_{[\alpha\beta\gamma]}=1/3!\left\{
A_{\alpha\beta\gamma}+A_{\beta\gamma\alpha}+A_{\gamma\alpha\beta}-
A_{\beta\alpha\gamma}-A_{\alpha\gamma\beta}-A_{\gamma\beta\alpha}\right\};$}
round brackets $(\dots)$ denote complete symmetrization:e.g.,

\centerline{$A_{(\alpha\beta\gamma)}=1/3!
\left\{A_{\alpha\beta\gamma}+A_{\beta\gamma\alpha}+A_{\gamma\alpha\beta}+
A_{\beta\alpha\gamma}+A_{\alpha\gamma\beta}+A_{\gamma\beta\alpha}\right\}.$}}.

The explicit expressions for the quantities in
(\ref{A16})  are given by the relations
\begin{eqnarray}
\label{A17}
P_1^{\mu\nu}&=&\left\{\td{p}{\!}^\mu{}_{\alpha}
-3(\td{x}k)\od{p}{\!}^\mu{}_{\alpha}+
3(\td{x}k)^2{p}^\mu{}_{\alpha}-
(\fd{x}k){p}^\mu{}_{\alpha}\right\}k^\alpha k^\nu, \\
\label{A18}
Q_1^{\mu\nu}&=&\od{p}{\!}^{\mu\nu}+3\od{p}{\!}^\mu_{\, \alpha}k^\alpha
k^\nu-4\od{p}{\!}^\mu{}_{\alpha}k^{(\alpha}\od{x}{}^{\nu)}-\nonumber\\
&-&(\td{x}k)p^{\mu\nu}-
6(\td{x}k)p^{\mu}{}_{\alpha}k^\alpha k^\nu+
6(\td{x}k)p^{\mu}{}_{\alpha}k^{(\alpha}
\od{x}{}^{\nu)}-2p^\mu{}_{\alpha}k^{(\alpha}\td{x}{}^{\nu)}\\
\label{A19}
R_1^{\mu\nu}&=&p^{\mu\nu}+3p^\mu{}_{\alpha}k^\alpha k^\nu-
6p^{\mu}{}_{\alpha}k^{(\alpha}\od{x}{}^{\nu)}+
2p^{\mu}{}_{\alpha}\od{x}{}^{\alpha}\od{x}{}^{\nu}.
\end{eqnarray}

In formulae  (\ref{A15}) and (\ref{A16}) the expressions
in the square brackets correspond to the contribution to the
$A_\mu$ and $F_{\mu\nu}$ of the charge of the particle,  while those in
the parentheses correspond to the contribution of the
dipole moment.

Inserting the explicit expression (\ref{A11}) for $p^{\mu\nu}$ in
(\ref{A15})-(\ref{A19}) we obtain the general form of the retarded
Lienard-Wiechert solution for the electromagnetic field strengths
of the charged spinning  particle in the pseudoclassical
theory.\\

{\bf3.} To disclose the correspondence between the obtained
pseudoclassical expressions for Lienard-Wiechert potentials and
well known classical formulae for the strengths of the electric
and magnetic fields of the charged particle with a magnetic
moment,  we will consider the case of the spinning particle at
rest ($\od{x}{\!}^i=0$,  $\od{x}{\!}^\mu =\delta_0^\mu $) with
precessing spin

From formulae (\ref{A16})-(\ref{A19}) we find the expression for the field
strengths of the electromagnetic field of the particle at rest:
\begin{eqnarray}
\label{A24}
&&F^{\mu\nu}_{\rm ret}=\left[\frac{g}{4\pi R^2} n^\mu\delta^\nu_0+
\frac{1}{4\pi R}\td {p}{}^{\mu\beta} n_\beta n^\nu+
\frac{1}{4\pi R^2}\left(\od {p}{\!}^{\mu\nu}+3\od
{p}{}^{\mu\beta} n_\beta n^\nu- 2\od {p}{}^{\mu\beta} n_\beta
\delta^\nu_0\right)+\right.\nonumber\\
&& +
\left. \frac{1}{4\pi R^3}\left( {p}^{\mu\nu}+
 3p^{\mu\beta} n_\beta n^\nu- 3 p^{\mu\beta}
 n_\beta\delta^\nu_0\right)\right]_{\tau=\tau_r}
  -\left[\frac{\!}{}\mu\leftrightarrow\nu\right]_{\tau=\tau_r},
\end{eqnarray}
where $n^\mu=R^\mu/R=(1, \vec{n})$,  $R=R_0=\sqrt{R_1^2+R_2^2+R_3^2}$
(for retarded fields $R_0=y_0-x_0(\tau)>0$).
The formula (\ref{A24}) was obtained  with taking into account the
fact that in the rest frame
$p^{0i}=0$,  which follows from (\ref{A23}). From
formula (\ref{A23}) at   $\vec{v}=0$ it also follows,  that
\begin{equation}
\label{A25}
p_{ij}=\frac{gM}{m^2}\varepsilon_{ijk0}W^k\od{x}{\!}^0=
-\frac{gM}{m^2}\varepsilon_{ijk}W_k
\end{equation}
($\varepsilon_{ijk}=-\varepsilon_{0ijk}=\varepsilon^{0ijk},
\varepsilon_{123}=1$).

Taking into account,  that in the rest frame of the particle we have
\cite{GG2}:
\begin{equation}
\label{A26}
          W_i=-mS_i,
\end{equation}
where $S_i$ is the spin of the particle,  we get
\begin{equation}
\label{A27}
p_{ij}=\frac{gM}{m}\varepsilon_{ijk}S_k.
\end{equation}

Hence $\od{p}_{ij}$ and $\td{p}_{ij}$ are equal to
\begin{equation}
\label{A28}
\od{p}_{ij}=\frac{gM}{m}\varepsilon_{ijk}\od{S}_k, \quad
\td{p}_{ij}=\frac{gM}{m}\varepsilon_{ijk}\td{S}_k.
\end{equation}
Let now the spin of the particle be precessing  around certain axis
with a constant angular velocity $\vec{\omega}$,   i.e.
\begin{equation}
\label{A29}
\od{S}_i=\left[{\vec{\omega}\vec{S}}\right]_i=\varepsilon_{ikl}\omega_
kS_l.
\end{equation}
Substituting  (\ref{A29}) into (\ref{A28}) we obtain
\begin{equation}
\label{A30}
\od{p}_{ij}=\frac{gM}{m}\left(\omega_iS_j-\omega_jS_i\right), \quad
\td{p}_{ij}=\frac{gM}{m}\left\{
\omega_i\left[{\vec{\omega}\vec{
S}}\right]_j-\omega_j\left[{\vec{\omega}\vec{S}}\right]_i
\right\}.
\end{equation}

Using (\ref{A27}),  (\ref{A30}) and also the equality $p^{0i}=0$,
we find from (\ref{A24}) the expressions for the
electric and magnetic field strengths for the charged particle
with precessing spin (the electric $E_i$ and magnetic $B_i$
fields are defined as $E_i\equiv(F^{01}, \, F^{02}, \, F^{03})$,
$B_i\equiv(F^{23}, \, F^{31}, \, F^{12})$):
\begin{eqnarray}
\label{A31}
E_i&=&-F_{0i}=\frac{g}{4\pi R^2}{n}_i- \nonumber\\
&-&\frac{gM}{4\pi mR}\left\{
\omega_i\left(\left[{\vec{\omega} \vec{S}}\right]\vec{n}\right)-
({\vec{\omega}\vec{ n})}\left[{\vec{\omega}
\vec{S}}\right]_i\right\}
-\frac{gM}{4\pi mR^2}\left\{
{\omega_i}({\vec{S}\vec{n}})-({\vec{\omega} \vec{n}})S_i\right\},
\end{eqnarray}
\begin{eqnarray}
\label{A32}
B_i=\frac{1}{2}\varepsilon_{ikl}F_{kl}&=&\frac{gM}{4\pi m
R^3}\left\{3\left({\vec{n}\vec{S}}\right)n_i-
S_i\right\}-\nonumber\\
&-&\frac{gM}{4\pi mR}\left\{
\left[\vec{\omega}\vec{n}\right]_i
\left(\left[{\vec{\omega}\vec{S}}\right]{
\vec{n}}\right)+
({\vec{\omega} \vec{n})}\left({\vec{n}
\vec{S}}\right)\omega_i-({\vec{\omega}
\vec{n}})^2S_i\right\}+\nonumber\\
&+&\frac{gM}{4\pi mR^2}\left\{
2\left[{\vec{\omega} \vec{S}}\right]_i-
3\left[{\vec{\omega} \vec{n}}\right]_i\left({\vec{n}
\vec{S}}\right)+
3({\vec{\omega} \vec{n})}\left[{\vec{S}\vec{n}}\right]_i\right\}.
\end{eqnarray}
From these formulae one can deduce the expressions for the
electric and magnetic fields of the particle at rest with a
fixed vector of the magnetic moment:
\begin{eqnarray}
\label{A33}
{E}_i&=&\frac{g}{4\pi R^2}{n}_i, \nonumber\\
{B}_i&=&\frac{gM}{4\pi mR^3}\left\{3({
\vec{n}\vec{S}}){n}_i-{S}_i\right\},
\end{eqnarray}
The relations (\ref{A31})-(\ref{A33})
coincide with the corresponding expressions for the particle
with a charge $g$ and total magnetic moment
$\displaystyle\vec{\mu}=\frac{gM}{m}\vec{S}$ \cite{BaT}.\\

{\bf 4.} Here we will obtain the formulae for the electromagnetic
field strengths of synchrotron radiation from  particles  in
circular orbits. We will consider  the  case  of  transverse
polarization of the particle ( when the spin of the particle
is perpendicular to the rotation  plane).  This  problem  is
interesting  from  the  point  of  view  of   measuring   the
polarization of the particles  by  the  measurement  of  the
synchrotron radiation.

It is well known,  that the radiation field is given by the terms
in the expression (\ref{A16}),  proportional to $1/\rho$:
\begin{equation}
\label{A34}
F_{{\rm rad}}^{\mu\nu}(y)=\left[\frac{g}{4\pi}\frac{2}{
\rho}\left(k^{[\mu}\td{x}{\!}^{\nu]}-
(\td{x}k)k^{[\mu}\od{x}{\!}^{\nu]}\right)
+\frac{1}{4\pi}\frac{2P_1^{\left[\mu\nu\right]}}{\rho}\right]_
{\tau=\tau_r},
\end{equation}
where $P_1^{\mu\nu}$ is given by formula (\ref{A17}).

Let us now  transform  the  formula  (\ref{A34})  taking  into
account the relations
\begin{eqnarray}
\label{A35}
\od{x}{\!}^\mu&=&\gamma(1,
\vec{v}), \quad\rho=\gamma(R-(\vec{v}\vec{R}))=\gamma
R(1-(\vec{v}\vec{n})),  \gamma=1/\sqrt{1-\vec{v}^2}\nonumber\\
\td{x}{\!}^\mu&=&\left(\gamma^4(\vec{v}\vec{a}), \, \, \gamma^2\vec{a}+
\gamma^4\vec{v}(\vec{v}\vec{a})\right), \quad
k^\alpha=\left(\frac{R}{\rho}, \frac{\vec{R}}{\rho}\right)=
\frac{R}{\rho}n^\alpha, \\
\fd{x}{\!}^\mu&=&\left(\od{a^0}, \, \, \gamma^3{\vec{b}}+
3\gamma^5 (\vec{v}\vec{a})\vec{a}+\od{a^0}\vec{v}\right), \quad
\od{a^0}=\gamma^5\left[(\vec{v}\vec{b})+\vec{a}^2\right]+
4\gamma^7(\vec{v}\vec{a})^2, \nonumber
\end{eqnarray}
where $\vec{v}\equiv (v^i)=dx^i/dt$ is the three-velocity of
the particle,
     $\vec{a}=d\vec{v}/dt$,            $\vec{b}=d\vec{a}/dt$,
$dt/d\tau=(1-\vec{v}^2)^{-1/2}=\gamma$.

Substituting (\ref{A35}) into the first term  of  expression
(\ref{A34})  for  $F^{\mu\nu}_{{\rm rad}}$  we  obtain  a   well   known
expression for  the  electric  field  strength  $E_{i}^{{\rm
ch}}$ of the charged spinless particle:
\begin{equation}
\label{A36}
E_{i}^{{\rm ch}}=\frac{g}{4\pi R(1-(\vec{v}\vec{n}))^3}
\left[\left(n_i-v_i\right)(\vec{a}\vec{n}) -
a_i(1-(\vec{v}\vec{n}))\right].
\end{equation}

To transform the spin part $P_1^{\mu\nu}$  of the tensor
$F^{{\mu\nu}}_{{\rm rad}}$,  which contains the $p^{\mu\nu}$,
connected with $W^\mu$  by the relation (\ref{A23}),
we will make use of the relation between  $W_\mu$  and  spin
vector of the particle in the rest frame $\vec{S}\equiv (S_i)$
\cite{GG2, GG5} and retaining only terms quadratic in $\xi^2$:
\begin{equation}
\label{A37}
\frac{W_\mu}{m}=\left(\frac{(\vec{{\cal P}^D}\vec{S})}{m},  \,  \,
{S}_i+\frac{{{\cal P}^D}_i(\vec{{\cal P}^D}\vec{S})}{m({\cal E}+m)}\right)\nonumber\\
=\left(\gamma(\vec{v}\vec{S}),  \,  \,  {S}_i+
\gamma^2\frac{{v}_i(\vec{v}\vec{S})}{\gamma+1}\right),
\end{equation}
where  ${\cal P}^D_\mu=-{\cal P}_\mu=({\cal E},  \vec{{\cal
P}^D})$ is the physical 4-momentum of the particle:
\begin{equation}
\label{A54}
({\cal P}^D_\mu)^2=m_{{\rm eff}}^2,  \qquad {\cal E}=\gamma{m_{{\rm eff}}},
\qquad \vec{{\cal P}^D}=\gamma{m_{{\rm eff}}}\vec{v},
\end{equation}
and $m_{{\rm eff}}$
is the effective mass of the particle,  which is defined by the
relation (\ref{A12A}).

Also from (\ref{A23}) with (\ref{A37}) taken into account we have
\begin{equation}
\label{A38}
p_{0i}=\frac{gM}{m^2}\varepsilon_{0ijk}W^j\od{x}{\!}^k=
-\gamma\frac{gM}{m}\varepsilon_{ijk}S_jv_k,
\end{equation}
\begin{equation}
\label{A39}
p_{ij}=\frac{gM}{m^2}\varepsilon_{ij\mu\nu}W^\mu\od{x}{\!}^\nu=
-\frac{gM}{m}\varepsilon_{ijk}\left[\gamma S_k-
\frac{\gamma^2}{\gamma+1}(\vec{v}\vec{S})v_k\right].
\end{equation}

In  the  case  of  the  uniform  circular  motion   of   the
transversely polarized particle  considered  here  (when  the
spin of the particle is along  the  vector  of  the  external
magnetic field $\vec{B}$) (see fig.1) the following relations hold:
\begin{eqnarray}
\label{A40}
&&(\vec{v}\vec{a})=(\vec{S}\vec{v})=(\vec{S}\vec{a})=
(\vec{S}{\vec{b}})=0, \nonumber \\
&& \od{\vec{S}}=0,  \quad
\od{\gamma}=0, \quad \vec{b}=-\omega^2 \vec{v},
\end{eqnarray}
where  $\omega$   is   the   circular   frequency   of   the
particle. With relations (\ref{A40}) taken into account  the
expression (\ref{A39}) takes the form
\begin{equation}
\label{A41}
p_{ij}=
-\frac{gM}{m}\gamma\varepsilon_{ijk}S_k.
\end{equation}

\hspace{1cm}
\unitlength=1mm
\linethickness{0.2mm}
\begin{picture}(55,  55)
\put(20,  25){\vector(1,  0){30}}
\put(20,  25){\vector(0,  1){30}}
\put(20,  25){\vector(-1,  -1){20}}
\put(20,  25){\vector(1,  2){10}}
\multiput(20,  25)(5,  -5){3}{\line(1,  -1){4}}
\multiput(30,  15)(0,  3){10}{\line(0,  1){2}}
\put(20,  35){\oval(10,  10)[tr]}
\put(26,  25){\oval(12,  12)[br]}
\put(4,  5){\shortstack{$\vec{B}(\vec{S})$}}
\put(23,  41){\shortstack{$\theta$}}
\put(16,  52){\shortstack{$\vec{v}$}}
\put(46,  27){\shortstack{$\vec{a}$}}
\put(31,  43){\shortstack{$\vec{n}$}}
\put(33,  21){\shortstack{$\varphi$}}
\put(23,  0){\shortstack{Fig.1}}
\end{picture}
\hspace{1cm}
\unitlength=1mm
\linethickness{0.2mm}
\begin{picture}(55,  55)
\put(20,  25){\vector(1,  0){30}}
\put(20,  25){\vector(0,  1){30}}
\put(20,  25){\vector(-1,  -1){20}}
\put(20,  25){\vector(1,  2){10}}
\multiput(20,  25)(5,  -5){3}{\line(1,  -1){4}}
\multiput(30,  15)(0,  3){10}{\line(0,  1){2}}
\put(22,  18){\oval(10,  20)[tr]}
\put(30,  25){\oval(20,  20)[br]}
\put(4,  5){\shortstack{$\vec{a}$}}
\put(26,  29){\shortstack{$\beta$}}
\put(46,  27){\shortstack{$\vec{v}$}}
\put(31,  43){\shortstack{$\vec{n}$}}
\put(40,  16){\shortstack{$\psi$}}
\put(10,  52){\shortstack{$\vec{B}(\vec{S})$}}
\put(23,  0){\shortstack{Fig.2}}
\end{picture}

From (\ref{A38}) and (\ref{A41}) using (\ref{A40}) we get
\begin{equation}
\label{A42}
\od{p}{\!}_{0i}=-\frac{gM}{m}\gamma^2\varepsilon_{ijk}S_ja_k, \quad
\td{p}{\!}_{0i}=-\frac{gM}{m}\gamma^3\varepsilon_{ijk}S_j{b}_k=
\frac{gM}{m}\omega^2\gamma^3\varepsilon_{ijk}S_j{v}_k, \quad
\od{p}{\!}^{ij}=\td{p}{\!}^{ij}=0.
\end{equation}

Transforming the second summand of the formula (\ref{A34}) for
$F^{\mu\nu}_{\rm rad}$ we find the expression for the
contribution of the dipole moment of the particle into the
radiation electric   field   strength $E_i$:
\begin{eqnarray}
\label{A43}
E_{i}^{\rm dip}&=&
-\frac{1}{4\pi \rho}\left(T_{0j}k^jk_i-T_{i\mu}k^\mu k_0\right)=
-\frac{R^2}{4\pi\rho^3}\left(T_{0j}\left(\delta_{ij}-n_in_j\right)+
T_{ij}n_j\right)\\
E_i^{\rm rad}&=&E_{i}^{{\rm ch}}+E_{i}^{{\rm dip}},\nonumber
\end{eqnarray}
where
 \begin{equation}
\label{A44}
T_{\mu\nu}=
\td{p}{\!}_{\mu\nu}
-3(\td{x}k)\od{p}{\!}_{\mu\nu}+
3(\td{x}k)^2{p}_{\mu\nu}-
(\fd{x}k){p}_{\mu\nu}.
\end{equation}
From  (\ref{A35}) with accounting for (\ref{A40}) we get
\begin{equation}
\label{A45}
\td{x}{\!}^\mu=\left(0, \, \, \gamma^2{a}^i\right), \quad
\fd{x}{\!}^\mu=\left(0, \, \, \gamma^3{b}^i\right)=
\left(0, \, \, -\gamma^3\omega^2
{v}^i\right).
\end{equation}
Now from (\ref{A44}) using (\ref{A38}),  (\ref{A41}),
(\ref{A42}) and (\ref{A45}) we obtain
\begin{equation}
\label{A46}
T_{0i}=-\frac{gM}{m}\gamma^3\varepsilon_{ijk}S_j\left[
\frac{3(\vec{a}\vec{n})a_k}{1-(\vec{n}\vec{v})}
+\frac{3v_k(\vec{a}\vec{n})^2}{(1-(\vec{n}\vec{v}))^2}-
\frac{\omega^2{v}_k}{1-(\vec{n}\vec{v})}
\right],
\end{equation}
\begin{equation}
\label{A47}
T_{ij}=-\frac{gM}{m}\gamma^3\varepsilon_{ijk}S_k\left[
\frac{3(\vec{a}\vec{n})^2}{(1-(\vec{n}\vec{v}))^2}-
\frac{\omega^2(\vec{n}\vec{v})}{1-(\vec{n}\vec{v})}
\right].
\end{equation}
In the case of  the  uniform  circular movement  of  the  transversely
polarized particle apart from relations (\ref{A40}) we  have
following evident relations:
\begin{equation}
\label{A48}
\varepsilon_{ijk}S_jv_k=\left[\vec{S}, \vec{v}\right]_i=
-\frac{S_B}{\omega}a_i, \quad
\varepsilon_{ijk}S_ja_k=\left[\vec{S}, \vec{a}\right]_i=
{S_B}{\omega}v_i,
\end{equation}
where $S_B$ is the projection of  the  spin  vector  on  the
direction of the magnetic field strength $\vec{B}$.
On account of these relations the expression (\ref{A46}) can
be rewritten as,
\begin{equation}
\label{A49}
T_{0i}=-\frac{gM}{m}\gamma^3\left[
\frac{\omega a_i}{1-(\vec{n}\vec{v})}+
\frac{3(\vec{a}\vec{n})\omega
v_i}{1-(\vec{n}\vec{v})}
-\frac{3(\vec{a}\vec{n})^2a_i}{\omega
(1-(\vec{n}\vec{v}))^2}\right]S_B.
\end{equation}

Substituting (\ref{A49}) and (\ref{A47}) into (\ref{A43})  we
find
\begin{eqnarray}
\label{A50}
E_i^{{\rm dip}}&=&\frac{gM}{4m\pi
R(1-(\vec{n}\vec{v}))^3}\left\{\left[
\frac{\omega a_j}{1-(\vec{n}\vec{v})}+
\frac{3(\vec{a}\vec{n})\omega
v_j}{1-(\vec{n}\vec{v})}
-\frac{3(\vec{a}\vec{n})^2a_j}{\omega
(1-(\vec{n}\vec{v}))^2}\right] S_B
(\delta_{ij}-n_in_j)+\right.\nonumber\\
&+&\left.\left(
\frac{3(\vec{a}\vec{n})^2}{(1-(\vec{n}\vec{v}))^2}-
\frac{\omega^2(\vec{n}\vec{v})}{1-(\vec{n}\vec{v})}
\right)\left[\vec{n}, \vec{S}\right]_i\right\}.
\end{eqnarray}
Using expressions (\ref{A36}) and (\ref{A50}),  and also  the
relation
\begin{equation}
\label{A51}
\left[\vec{n}, \vec{S}\right]_i=\frac{S_B}{\omega
v^2}\left((\vec{v}\vec{n})a_i -(\vec{a}\vec{n})v_i\right),
\end{equation}
which is obtained using expressions (\ref{A40}) and (\ref{A48}),
we can write the expression for  the  synchrotron  radiation
rate of the particle per unit solid angle $d\Omega$ and per unit
time of the "particle's own time" (at the moment of radiation)
$t_0$:
\begin{equation}
\label{A52}
\frac{d  I}{dt_0 d\Omega}=\left(\frac{g}{4\pi}\right)^2
\frac{a^2}{(1-v\cos\theta)^5}\left[
(1-v\cos\theta)^2-(1-v^2)\sin^2\theta\cos^2\varphi-
2\frac{M}{m}\omega S_B\sin^2\theta\sin^2\varphi\right],
\end{equation}
where $\theta$ and $\varphi$ are  the  angles  defining  the
orientation  of  the  vector  $\vec{n}$  in  the   cartesian
coordinate system defined by three vectors
$\vec{v},  \vec{a}$ and $\vec{B}(\vec{S})$
(see fig.1).

Integrating the expression (\ref{A52}) over the solid angle
we get for the radiation rate the formula:
\begin{equation}
\label{A53}
\frac{d I}{dt_0
}=\frac{g^2}{4\pi}\frac{2a^2}{3(1-v^2)^2}\left[1-
\frac{M}{m}\frac{\omega S_B}{1-v^2}\right].
\end{equation}
To compare this with the formulae obtained in quantum radiation theory,
note,  that the effective mass of the particle,  defined by the
expression  (\ref{A12A}),  in the case of motion in the
homogeneous external magnetic field takes the form
\begin{equation}
\label{A55}
m_{{\rm
eff}}=m+\frac{i}{2}\frac{gM}{m}F_{ij}\xi_i\xi_j=
m+\frac{1}{2}F_{ij}p_{ij}.
\end{equation}
The right-hand side of the formula (\ref{A55}) is written on
account  of  the  relation   (\ref{A11}) and the gauge
(\ref{A20}).   Substituting   in (\ref{A55})  the expression for
$p_{ij}$ from (\ref{A41}) we find for the $m_{{\rm eff}}$ (to the
first order in spin)
\begin{equation}
\label{A56}
m_{{\rm eff}}=m-\frac{{\cal E}}{m}\frac{gM}{m} B_iS_i.
\end{equation}
Using (\ref{A56}) we obtain from (\ref{A54})
 to the first order in spin  the relation
\begin{equation}
\label{A57}
\frac{1}{(1-v^2)^n}=\left(\frac{{\cal E}}{m}\right)^{2n}
(1+2n\frac{gM}{m^2}\frac{{\cal E}}{m} BS_B)=
\left(\frac{{\cal E}}{m}\right)^{2n}
(1+2n\frac{M}{m}\left(\frac{{\cal E}}{m}\right)^2S_B\omega).
\end{equation}
In deducing the last equation in (\ref{A57}) we used the relation
$gB=m\omega\gamma$. Using now the equality  (\ref{A57}),   the
formula  (\ref{A53})   can   be rewritten in the form
\begin{equation}
\label{A58}
\frac{d I}{dt_0}=\frac{g^2}{4\pi}\frac{2a^2}{3}
\left(\frac{{\cal E}}{m}\right)^4\left[1
+3\frac{M}{m}S_B\omega\left(\frac{{\cal E}}{m}\right)^2\right].
\end{equation}
On account of the relation $S=\frac{1}{2}\zeta$,  where
$\zeta=\pm1$ is the projection of the polarization vector on
the direction of  the  magnetic  field,   we  find  that  this
formula (\ref{A58}) coincides with that found in the quantum theory 
of radiation (to the first order in spin).

To investigate the characteristics of the polarization of
synchrotron radiation we resolve (following \cite{BKF}) the electric field
strength $\vec{E}^{\rm rad}$ into components along the vectors
$\vec{e}=\vec{a}/a$  and  $\left[\vec{n}, \vec{e}\right]$:
\begin{equation}
\label{A59A}
\vec{E}^{\rm rad}=E_1 \vec{e}+E_2\left[\vec{n}, \vec{e}\right],
\end{equation}
This choice of unit vectors is suitable for description of
the radiation of the relativistic particles using angles defined
on Fig.2,  which are appropriate since the main contribution to
the radiation comes from small $(\sim 1/\gamma)$ angles $\beta$
and $\psi$,  in accordance with the fact,  that the angle between
vectors $\vec{n}$ and $\vec{v}$ is of the order of $(\sim
1/\gamma)$. As it was mentioned in \cite{BKF},  the vector
$\vec{E}$ is not orthogonal to vector $\vec{n}$,  however this
deviation is of the order of $1/\gamma$,  so that the
decomposition (\ref{A59A}) can be used for calculation of major
terms (with accuracy $1/\gamma$).
From formulae (\ref{A36}) and  (\ref{A50}) we get
according to decomposition (\ref{A59A}) the expressions for
$E_1$  and $E_2$:
\begin{equation}
\label{A59}
E_1=\frac{ga}{\pi
R(\mu^2+\psi^2)^3}\left((\psi^2-\mu^2)+\frac{M}{m}S_B\omega
\frac{4\beta^2(\mu^2-5\psi^2)}{(\mu^2+\psi^2)^2}
\right)_{\tau=\tau_r},
\end{equation}
\begin{equation}
\label{A60}
E_2=-\frac{2ga}{\pi
R(\mu^2+\psi^2)^3}\left(\beta\psi-\frac{M}{m}S_B\omega
\frac{4\beta\psi(2\mu^2-\psi^2)}{(\mu^2+\psi^2)^2}
\right)_{\tau=\tau_r},
\end{equation}
where $\mu^2=\gamma^{-2}+\beta^2$.

Taking into account the equality (\ref{A57}),  we  rewrite the
expression for  $\mu^2$ in the form
\begin{equation}
\label{A61}
\mu^2=\beta^2+\left(\frac{m}{{\cal
E}}\right)^2-2\frac{M}{m}S_B\omega=\mu_0^2-2\frac{M}{m}S_B\omega, \quad
\mu_0^2=\beta^2+\left(\frac{m}{{\cal E}}\right)^2.
\end{equation}

Substituting now the expression for $\mu^2$ into (\ref{A59}) and
(\ref{A60}),  we get to the first order in spin
\begin{equation}
\label{A62}
E_1=\frac{ga}{\pi
R(\mu^2_0+\psi^2)^3}\left[(\psi^2-\mu^2_0)+4\frac{M}{m}S_B\omega
\left(\frac{2\psi^2-\mu_0^2}{\mu_0^2+\psi^2}+
\frac{\beta^2(\mu_0^2-5\psi^2)}{(\mu^2_0+\psi^2)^2}
\right)\right]_{\tau=\tau_r},
\end{equation}
\begin{equation}
\label{A63}
E_2=-\frac{2ga\beta\psi}{\pi
R(\mu^2_0+\psi^2)^3}\left(1-\frac{M}{m}S_B\omega
\frac{2(\mu^2_0-5\psi^2)}{(\mu^2_0+\psi^2)^2}
\right)_{\tau=\tau_r}.
\end{equation}

For $\mu_0^2-\psi^2=0$ from  (\ref{A62}) and (\ref{A63}) follows,
that
\begin{equation}
\label{A64}
E_1=\frac{ga}{4\pi
R\mu^6_0}\frac{M}{m}S_B\omega\left(1-\frac{2\beta^2}{\mu_0^2}\right)
_{\tau=\tau_r},
\end{equation}
\begin{equation}
\label{A65}
E_2=-\frac{ga\beta\psi}{4\pi
R\mu^6_0}\left(1+\frac{2MS_B\omega}
{m\mu^2_0}
\right)_{\tau=\tau_r}.
\end{equation}
From obtained formulae one can see,  that in the direction
$\psi=m/{\cal E}$ of the plane $(\vec{a}\vec{v})$
($\beta=0$) there is a radiation
(proportional to the spin),  contrary to the case of particle
without spin.\\

Authors are thankful to I.V.Tyutin and A.Airapetian for useful
discussion of the topics of the paper.

This investigation was supported in part by the grants
INTAS 96-538,  INTAS 93-1038 и INTAS-RFBR 95-0829.

\end{document}